\documentclass[10pt, twocolumn, pre, aps, superscriptaddress, showpacs]{revtex4-1}
\usepackage{amsmath, graphicx, subfigure, tikz}
\begin{document}

\title{The Merged Potts-Clock Model: Algebraic and Conventional\\
Multistructured Multicritical Orderings in Two and Three Dimensions}
\author{E. Can Artun}
    \affiliation{T\"UBITAK Research Institute for Fundamental Sciences, Gebze, Kocaeli 41470, Turkey}
    \affiliation{Faculty of Engineering and Natural Sciences, Kadir Has University, Cibali, Istanbul 34083, Turkey}
\author{A. Nihat Berker}
    \affiliation{Faculty of Engineering and Natural Sciences, Kadir Has University, Cibali, Istanbul 34083, Turkey}
    \affiliation{T\"UBITAK Research Institute for Fundamental Sciences, Gebze, Kocaeli 41470, Turkey}
    \affiliation{Department of Physics, Massachusetts Institute of Technology, Cambridge, Massachusetts 02139, USA}

\begin{abstract}
A spin system is studied, with simultaneous permutation-symmetric Potts and spin-rotation-symmetric clock interactions, in spatial dimensions $d=2$ and 3.  The global phase diagram is calculated from the renormalizaton-group solution with the recently improved (spontaneous first-order detecting) Migdal-Kadanoff approximation or, equivalently, with hierarchical lattices with the inclusion of effective vacancies.  Five different ordered phases are found: conventionally ordered ferromagnetic, quadrupolar, antiferromagnetic phases and algebraically ordered antiferromagnetic, antiquadrupolar phases. These five different ordered phases and the disordered phase are mutually bounded by first- and second-order phase transitions, themselves delimited by multicritical points: inverted bicritical, zero-temperature bicritical, tricritical, second-order bifurcation, and zero-temperature highly degenerate multicritical points. One rich phase diagram topology exhibits all of these phenomena.
\end{abstract}
\maketitle

\section{Introduction: Two Models Merged}

The $q$-state Potts models, ever since the establishment of their quantitative relevance to surface phase transitions \cite{BOP} and of the intricate renormalization-group mechanism for their changeover from second- to first-order phase transitions \cite{spinS7,AndelmanPotts1}, have held high interest in statistical physics.  The $q$-state clock models, ever since the establishment of their algebraic ordering in relation to the XY model \cite{Jose}, have also held high interest.  In the current work, we merge the two models into the $q$-state Potts-clock models and solve, in spatial dimensions $d=2$ and $d=3$, with the recently improved Migdal-Kadanoff approximation \cite{Devre} or, equivalenty, exactly on hierarchical lattices \cite{BerkerOstlund,Kaufman1,Kaufman2,BerkerMcKay}, obtaining algebraically \cite{BerkerKadanoff1,BerkerKadanoff2,Saleur,Ilker3} and conventionally ordered multistructured multicitical global phase diagrams. This merged model has been recently studied \cite{Polackova} on the square lattice for $q=6$ for positive couplings $J$ and $K$ (see below) by the corner transfer matrix renormalization-group method, showing the ferromagnetic phase with a single phase transition from the disordered phase for the Potts limit and with a narrow intermediate BKT phase in the clock limit.

The Potts and clock models, by themselves, have a lower-critical dimension above which a low-temperature ordered phase occurs.  In the antiferromagnetic case, the low-temperature phase is algebraically ordered when ground-state entropy occurs, as in all Potts models and the clock models with an odd number of states $q$.  For ferromagnetic Potts models, the phase transitions are second order only for low $d$ and low $q$.  All of these properties are obtained by position-space renormalization-group methods, as used in this study.\cite{spinS7,AndelmanPotts1,Devre,BerkerKadanoff1,BerkerKadanoff2,Ilker3}  For the ferromagnetic $q=5$ clock model on the square lattice, the phase transition occurs with a narrow intermediate BKT phase. \cite{Polackova}

The merged model is defined by the Hamiltonian
\begin{equation}
- \beta {\cal H} = \sum_{\left<ij\right>} \, [J\, \delta(\vec s_i,\vec s_j)  \, + K \, \vec s_i \cdot \vec s_j],
\end{equation}
where $\beta=1/k_{B}T$, at site $i$ the spin $\vec s_i$ can point in $q$ different directions $\theta _i = 2\pi n_i/q$ in the $xy$ plane, with $n_{i}=0,1,...,q-1$ providing the $q$ different possible states, the delta function $\delta(\vec s_i,\vec s_j)=1(0)$ for $\vec s_i = \vec s_j (\vec s_i \neq \vec s_j)$, and the sum is over all interacting pairs of spins. We independently vary the Potts interaction strength $J$ and the clock interaction strength $K$ of the merged Potts-clock model, to obtain the multistructured multicritical global phase diagram.

\begin{figure}[ht!]
\centering
\includegraphics[scale=0.40]{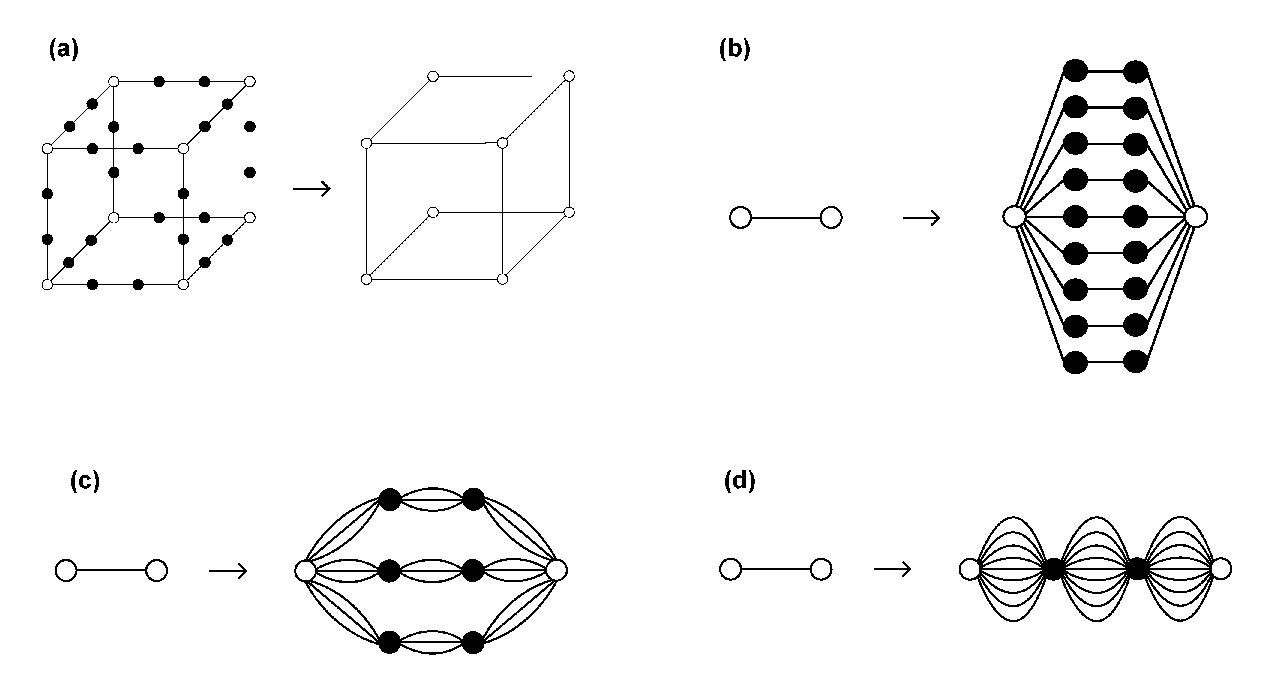}
\caption{(a)  The Migdal-Kadanoff approximate renormalization-group transformation on the cubic lattice. Bonds are removed from the cubic lattice to make the renormalization-group transformation doable.  The removed bonds are compensated by adding them to the remaining bonds: before, after, or partially before partially after. In each of (b-d), a hierarchical model is constructed by self-imbedding a graph into each of its bonds, \textit{ad infinitum}.\cite{BerkerOstlund}  The exact renormalization-group solution proceeds in the reverse direction, by summing over the internal spins shown with the dark circles.  Shown in (b-d) are the most used, so called "diamond" hierarchical lattices \cite{BerkerOstlund,Kaufman1,Kaufman2}.  The length-rescaling factor $b$ is the number of bonds in the shortest path between the external spins shown with the open circles, $b=3$ in these cases. The volume rescaling factor $b^d$ is the number of bonds replaced by a single bond, $b^d=27$ in these cases, so that $d=3$.  In the renormalization-group solutions, in (b), $b^{d-1}$ bond moving is done after summing over (namely decimating) the internal spins along a length-rescaling line of $b$ bonds.  In (d), $b^{d-1}$ bond moving is done before decimation over $b$ bonds.  In (c), the fraction $f$ of the bond moving is done before and the remaining fraction $1-f$ is done after the decimation.}
\end{figure}

\section{Method: Migdal-Kadanoff Approximation, Improved, and Hierarchical Lattices}
\begin{figure*}[ht!]
\centering
\includegraphics[scale=0.7]{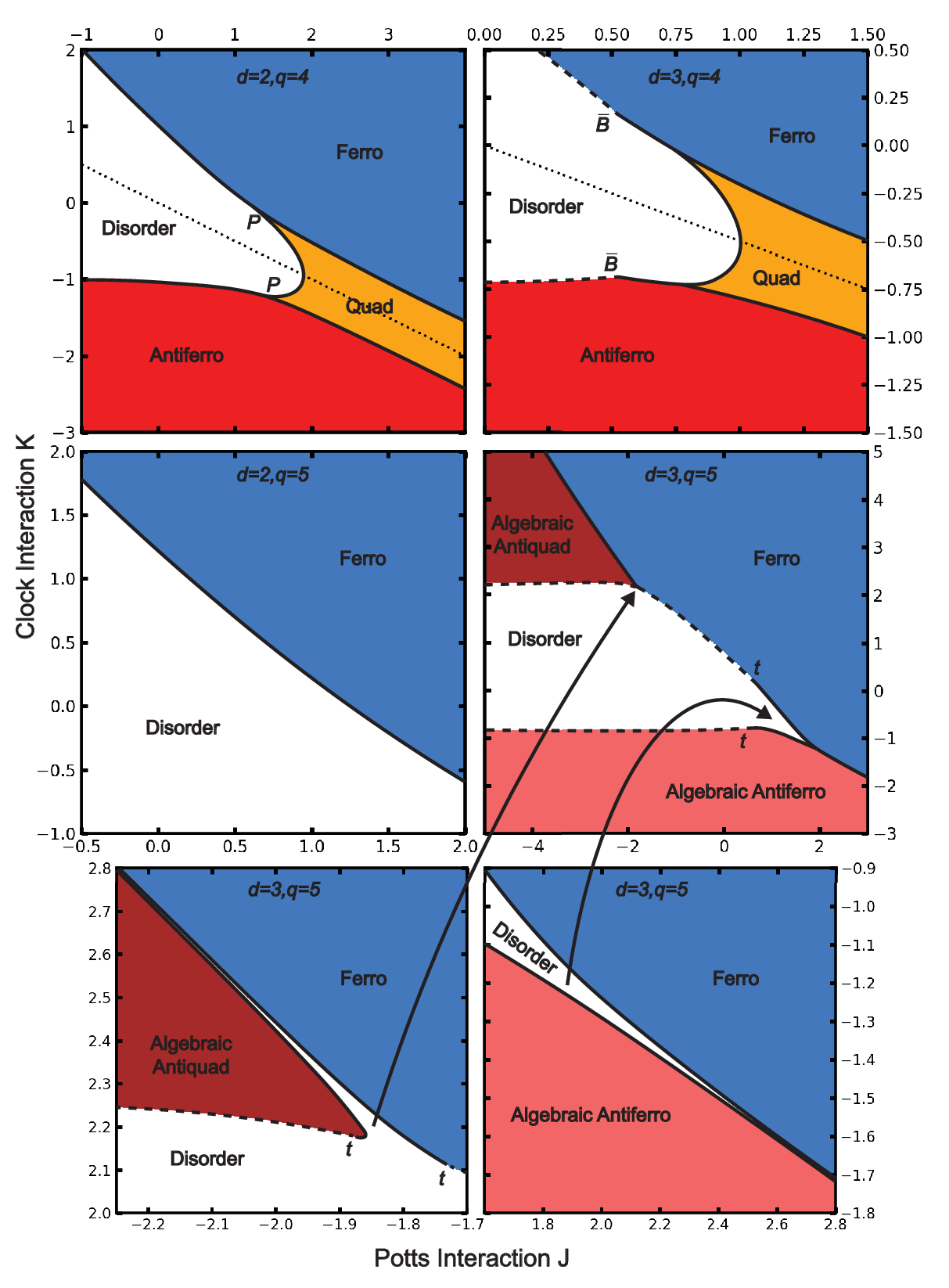}
\caption{Calculated phase diagrams of the $q=4$ and 5 state Potts-clock models in $d=2$ and 3, in terms of the respective interactions $J$ and $K$.  First- and second-order phase transition lines are shown with dashed and full lines, respectively.  The disorder line, occurring in $q=4$, is shown with a dotted line. Two types (namely, bordered by different phases) of inverted (see Fig. 3) bicritical points $\overline{B}$, two types of second-order bifurcation points $P$, three types of tricritical points $t$ are seen.  As seen in the insets (bottom two panes), the narrow disordered phase, between the algebraically ordered ferromagnetic phase or the algebraically ordered quadrupolar phase and the ferromagnetic phase, narrowly extends to two types of zero-temperature (see insets) bicritical points $Z$ (Fig. 3). We recall that our calculation is a numerically exact solution of the models on a hierarchical lattice, so that the phase boundaries are obtained beyond the accuracies of the thickness of the lines in the figures, as seen in the insets here. Both in $d=2$ and 3, the conventionally ordered quadrupolar phase $(q=4$) non-narrowly extends to zero-temperature highly degenerate multicritical point $S$, given in Fig. 3. Thus, a qualitatively very different picture emerges for even and odd number of states $q$.}
\end{figure*}
\begin{figure*}[ht!]
\centering
\includegraphics[scale=0.7]{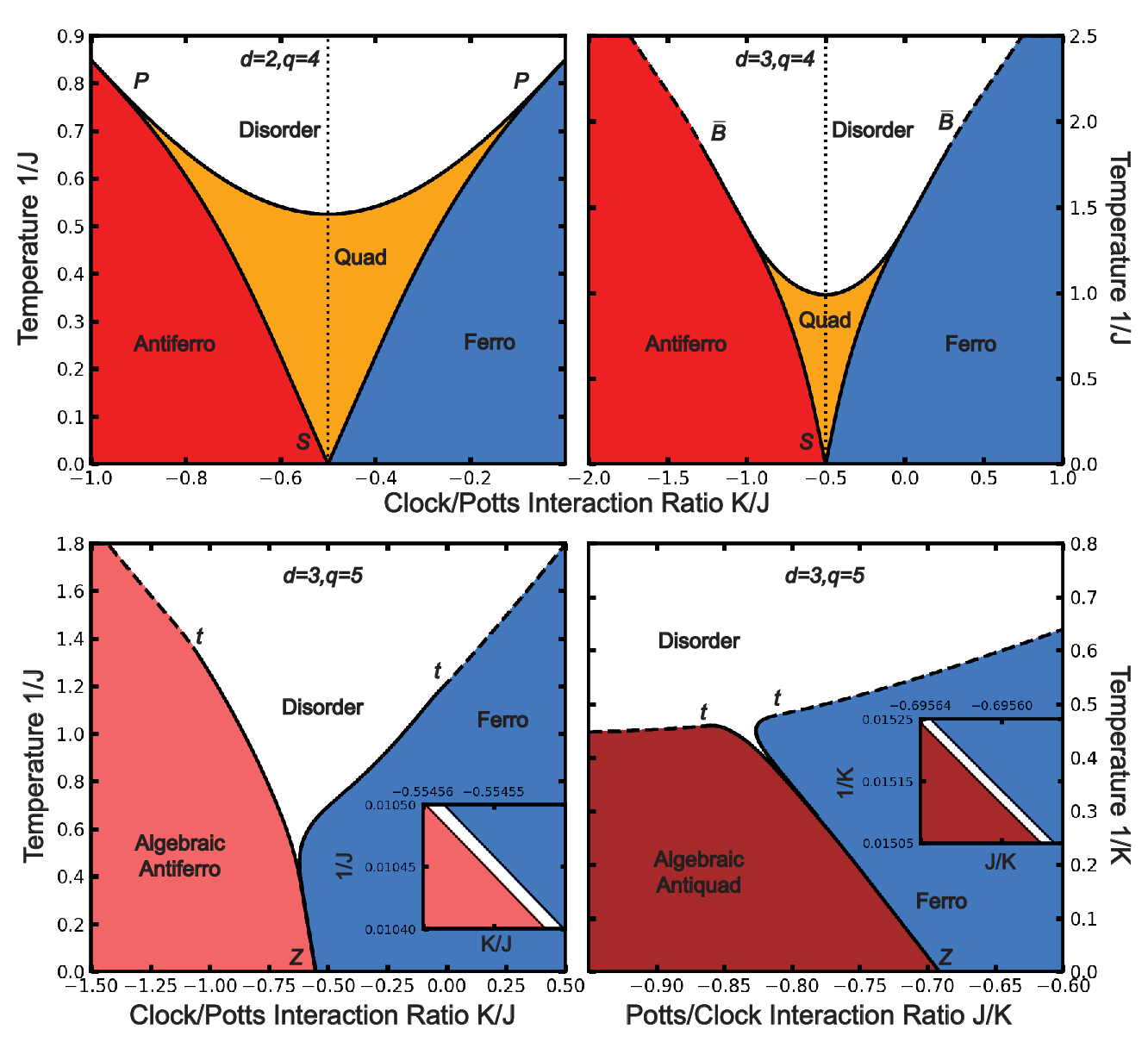}
\caption{Calculated strong-coupling behaviors in $d=2$ and $d=3$, in terms of the temperature variables $1/J$ or $1/K$ and the ratio of the Potts-clock interactions $K/J$ or $J/K$. First- and second-order phase transitions are shown with dashed and full lines, respectively. Top row: The $q=4$ state Potts-clock models.  The disorder line is shown with a dotted line. Both in $d=2$ and 3, the conventionally ordered quadrupolar phase non-narrowly extends to the zero-temperature highly degenerate multicritical point $S$. Bottom row: The $q=5$ state Potts-clock model in $d=3$. As also seen in the insets, the narrow disordered phase, between the algebraically ordered ferromagnetic phase or the algebraically ordered quadrupolar phase and the ferromagnetic phase, narrowly extends to two types (namely, bordered by different phases) of zero-temperature (see insets of Fig. 2) bicritical points $Z$.  As also seen in Ref. \cite{BerkerKadanoff1,BerkerKadanoff2}, the algebraically ordered phases do not occur in $d=2$.  Two different types of tricritical points $t$ occur in each pane of the figure. }
\end{figure*}

The Migdal-Kadanoff approximation \cite{Migdal,Kadanoff} renders a non-doable renormalization-group transformation doable by a physically motivated approximate step, is very easily calculated, flexibly applicable to large number of systems, highly used and highly successful.  For example (Fig. 1a), an exact renormalization-group transformation cannot be applied to the cubic lattice.  Thus, as an approximation, some of the bonds are removed.  However, this weakens the connectivity of the system and, to compensate, for every bond removed, a bond is added to the remaining bonds.  This whole step is called the bond-moving step and constitutes the approximate step of the renormalization-group transformation.  At this point, the intermediate sites can be eliminated by an exact summation over their spin values in the partition function, which yields the renormalized interaction between the remaining sites.  This is called the (exact) decimation step and completes the renormalization-group transformation.

As acceptable as the procedure just described, the removed bonds can be compensated by adding the appropriate number of bonds to the result of the decimation, this being the number of decimated bonds that the removed bonds would have given.  Alternately, a certain fraction $f$ of the removed bonds could be compensated before decimation and the remaining fraction $1-f$ could be compensated after the decimation.  The choice of $0\leq f\leq 1$ is left to us.

Furthermore, as shown in Figs. 1b-c, the renormalization-group recursion relations of the Migdal-Kadanoff approximation are identical to those of an exactly solved hierarchical model \cite{BerkerOstlund,Kaufman1,Kaufman2,BerkerMcKay}, making the Migdal-Kadanoff approximation a physically realizable approximation, as used in polymers, electronic systems, and turbulence, respectively in Refs. \cite{Flory,Kaufman,Lloyd,Kraichnan}, and therefore a robust approximation. Hierarchical models \cite{BerkerOstlund,Kaufman1,Kaufman2,BerkerMcKay} are exactly solvable microscopic models that are currently widely used.\cite{Clark,Kotorowicz,ZhangQiao,Jiang,Chio,Myshlyavtsev,Derevyagin,Shrock,Monthus,Sariyer}   The construction of hierarchical models is illustrated in Fig. 1. Each line segment in Fig. 1 represents a nearest-neighbor spin-spin interaction $J\,\delta(\vec s_i,\vec s_j)  \, + K \,\cos(\vec s_i \cdot \vec s_j)$ as given in Eq.(1). In each of (b-d), a hierarchical model is constructed by self-imbedding a graph into each of its bonds, \textit{ad infinitum}.\cite{BerkerOstlund} Figs. 1b-c show hierarchical lattices for bond-moving before (Fig. 1b), after (Fig. 1d), or a combination as explained above (Fig. 1c). The exact renormalization-group solution proceeds in the reverse direction, by summing over the internal spins shown with the dark circles.

In the current study, our calculation corresponds to the hierarchical model in Fig.1(c), with the factor $f$ chosen so that our calculation yields the exact transition temperature of the model with $q=2$, namely the Ising model. This choice was used previously, e.g., in the quantum mechanical renormalization-group study of high-temperature superconductivity in the tJ model of electronic conduction.\cite{Falicov,highTc}  Thus, in the current study, the exact critical temperatures \cite{Onsager,Talapov} of $1/(J/2+K)=2.26918531$ and 4.51152785 are obtained, in $d=2$ and 3, with $f=0.5459793$ and 0.1775492, respectively.  Note that for $q=2$, both the Potts and clock terms in Eq.(1) reduce to the Ising model, with combined interaction constant $J/2+K$.

The above can be rendered algebraically in the most straightforward way by writing the transfer matrix between two neighboring spins, for example for $q=4$,
\begin{multline}
\textbf{T}_{ij} \equiv e^{-\beta {\cal H}_{ij}} =
\left(
\begin{array}{cccc}
e^{J+K} & 1 & e^{-K} & 1 \\
1 & e^{J+K} & 1 & e^{-K} \\
e^{-K} & 1 & e^{J+K} & 1 \\
1 & e^{-K} & 1 & e^{J+K}\end{array} \right),
\end{multline}
and for $q=5$,
\begin{multline}
\textbf{T}_{ij} \equiv e^{-\beta {\cal H}_{ij}} = \\
\left(
\begin{array}{ccccc}
e^{J+K} & e^{0.31K} & e^{-0.81K} & e^{-0.81K} & e^{0.31K} \\
e^{0.31K} & e^{J+K} & e^{0.31K} & e^{-0.81K} & e^{-0.81K} \\
e^{-0.81K} & e^{0.31K} & e^{J+K} & e^{0.31K} & e^{-0.81K} \\
e^{-0.81K} & e^{-0.81K} & e^{0.31K} & e^{J+K} & e^{0.31K} \\
e^{0.31K} & e^{-0.81K} & e^{-0.81K} & e^{0.31K} & e^{J+K}\end{array} \right),
\end{multline}
where $-\beta {\cal H}_{ij}$ is the part of the Hamiltonian between the two spins at the neighboring sites $i$ and $j$. An important degeneracy difference between these two transfer matrices, with important phase diagram consequences, will be discussed below.

The bond-moving step of the Migdal-Kadanoff approximate renormalization-group transformation consists in taking, before decimation, the power of $fb^{d-1}$ of each element in this matrix and in taking, after decimation, the power of $(1-f)b^{d-1}$ of each element in this matrix.  Here $b$ is the length-rescaling factor of the renormalization-group transformation, namely the renormalized nearest-neighbor separation in units of unrenormalized nearest-neighbor separation.  The decimation step consists in matrix-multiplying $b$ transfer matrices.  The flows, under this transformation, of the transfer matrices determine the phases, the phase transitions and all of the thermodynamic densities of the system, as illustrated below.

An important aspect of an occurring phase transition is the order of the phase transition. The $q$-state Potts models have a second-order phase transition for $q\leq q_c$ and a first-order phase transition for $q>q_c$.\cite{Baxter,duality1,MonteCarlo} In renormalization-group theory, this has been understood and reproduced as a condensation of effective vacancies formed by regions of disorder.\cite{spinS7,AndelmanPotts2}
The above has been included \cite{Devre} as a local disorder state into the two-spin transfer matrix of Eq.(2).  Inside an ordered region of a given spin value, a disordered site does not significantly contribute to the energy in Eq.(1), but has a multiplicity of $q-1$.  The substraction comes from the fact that the disordered site cannot be in the spin state of its surrounding ordered region.  This is equivalent to the exponential of an on-site energy and, with no approximation, is shared on the transfer matrices of the $2d$ incoming bonds.  The transfer matrix becomes, for example for $q=4$,

\begin{widetext}
\begin{gather}
\textbf{T}_{ij} \equiv e^{-\beta {\cal H}_{ij}} =
\left(
\begin{array}{ccccc}
e^{J+K} & 1 & e^{-K} & 1 & (q-1)^{1/2d} \\
1 & e^{J+K} & 1 & e^{-K} & (q-1)^{1/2d} \\
e^{-K} & 1 & e^{J+K} & 1 & (q-1)^{1/2d} \\
1 & e^{-K} & 1 & e^{J+K} & (q-1)^{1/2d} \\
(q-1)^{1/2d} & (q-1)^{1/2d} &(q-1)^{1/2d} & (q-1)^{1/2d} &(q-1)^{1/d}\end{array} \right).
\end{gather}
\end{widetext}

\section{Global Phase Diagrams}
Phase diagram predictions can be made from the \textit{a priori} examination of the Hamiltonian of the Potts-clock model in Eq.(1).  For the $q=4$ model (and in general for all even $q$ Potts-clock models), for interaction ratio $J/K=-2$, a cancellation occurs between the Potts and clock terms and the energies are equal for the completely aligned $(n_i=n_j)$ and completely antialigned $(|n_i-n_j|=q/2)$ interacting pairs of spins.  Thus, along this line on the phase diagram, all phases must be invariant under $\pi$ rotation of any individual spin.  In fact, only the quadrupolar \cite{Sivardiere} and disordered phases are seen along this line in our calculated phase diagrams (Figs. 2,3).  Indeed, in the quadrupolar phase, the neighboring spins are, randomly, either aligned or $\pi$-antialigned.  For the $q=5$ model (and in general for all odd $q$ Potts-clock models), the energies are equal for the two most antialigned (but cannot be completely antialigned due to odd $q$) pairs of spins $(|n_i-n_j|\pm 1/2=q/2)$, so that for interactions favoring antialignment, there is a ground-state energy degeneracy.  Thus, fluctuations will occur no matter how low the temperature, leading to a nonzero-temperature sink fixed point if an ordered phase occurs, making the latter algebraically ordered.\cite{BerkerKadanoff1,BerkerKadanoff2}  This is in fact what is seen, with the algebraic antiferromagnetic and algebraic antiquadrupolar phases in our calculated phase diagrams (Figs. 2,3).

Under repeated renormalization-group transformations, the phase diagram points of the ordered phases of the Potts-clock model flow to the sinks shown in Table I.  The sink values of the transfer matrix elements epitomize the whole basin of attraction of the completely stable fixed point that is the sink.  For example, in the ferromagnetic phase the spins are aligned along one of the $q$ spin directions, in the antiferromagnetic phase the spins up-down alternate along a spatial direction, in the quadrupolar phase the spins align, randomly, in a spin direction and its opposite direction. The algebraically ordered phases are discussed further below.  The disordered phase has two sinks, one sink with the lower-right $(q+1)\times (q+1)$ element of the transfer matrix equal to 1 and the rest zero, another sink with all elements in the upper-left $q\times q$ block equal to one and the rest zero.  Analysis at the unstable fixed points attracting the phase boundaries give the order of the phase transition.\cite{BOP} Our calculated phase diagrams are shown in Figs. 2,3.
\begin{table*}
\begin{tabular}{c c c c c c c}

\hline

\vline & & & Sinks of the Ordered Phases of the (q=4)-state Potts-Clock Model & &  &\vline  \\
\hline
\vline & [0,0,1,0] &\vline  & [1,0,1,0]  &\vline  & [1,0,0,0] &\vline   \\
\hline
\vline & Antiferromagnetic  &\vline & Quadrupolar  &\vline & Ferromagnetic &\vline \\
\hline

\vline & & & Sinks of the Ordered Phases of the (q=5)-state Potts-Clock Model & &  &\vline  \\
\hline
\vline & [0,1/3,1,1,1/3] &\vline  & [0,1,1/3,1/3,1]  &\vline & [1,0,0,0,0] &\vline   \\
\hline
\vline & Algebraic Antiferromagnetic  &\vline & Algebraic Antiquadrupolar  &\vline & Ferromagnetic&\vline  \\
\hline

\end{tabular}
\caption{Under repeated renormalization-group transformations, the phase diagram points of the ordered phases of the Potts-clock model flow to the sinks shown on this Table.  Only the top row of the sink transfer matrix is shown here.  The subsequent rows are obtained by cyclically rotating the elements.  The last row and last column of the transfer matrix, corresponding to the effective vacancies, have all elements zero at the ordered sinks and are not given here.}
\end{table*}

For $q=4$ (Figs. 2,3), ferromagnetic and antiferromagnetic phases, with intervening quadrupolar \cite{Sivardiere} and disordered phases, are seen.  The quadrupolar phase intervenes between the ferromagnetic and antiferromagnetic phases, up to a second-order bifurcation point $P$ in $d=2$ and up to an inverted bicritical \cite{Nelson,Hoston} point $\overline{B}$ in $d=3$.  The bicritical points are inverted, namely, their first-order stem is on the high-temperature side (Fig. 3), whereas in previously studied bicritical points the first-order stem extends towards low temperature.  All other phase transitions are second order.  A highly degenerate multicritical point $S$ occurs at $1/J=0,K/J=-0.5$ (Fig. 3), due to the degeneracy discussed at the beginning of this Section.  The ferromagnetic, quadrupolar, antiferromagnetic phases meet at this single zero-temperature multicritical point \cite{Hoston} in both $d=2$ and 3.

For $q=5$ (Figs. 2,3), in $d=3$, an algebraically ordered antiferromagnetic phase or the algebraically ordered antiquadrupolar phase and the ferromagnetic phase are separated by a narrow disordered phase terminating at a zero-temperature bicritical point.  In both cases, at the phase boundaries at higher temperatures on each side of the phase diagram, a tricritical point separates first- and second-order transition lines (Fig. 3).  In $d=2$, second-order phase transitions separate the ferromagnetic and disordered phases. It is seen from Table I that the sinks of the $q=5$ antiferromagnetic and antiquadrupolar phases have a temperature scale, namely that all elements of the sink transfer matrix are not 1 or 0.  In general, at a renormalization-group fixed point, the system is scale-invariant, so that the correlation length $\xi$ is zero (at disordered or conventionally ordered phase sinks), which cannot be if there is a temperature scale, or infinity (at fixed points attracting critical systems).\cite{Wilson1,Wilson2} Thus, in the present case the entirety of these antiferromagnetic and antiquadrupolar phases are critical.\cite{BerkerKadanoff1,BerkerKadanoff2,Saleur}.  The correlation length $\xi$ is infinite throughout these phases and, having no length scale, the phases are algebraically ordered.  In the $d=2$, the algebraically ordered phases are not seen.  Previous work has consistently shown for both Potts \cite{BerkerKadanoff1,BerkerKadanoff2} and odd-$q$ clock \cite{Ilker3} models, that the algebraically ordered phases occur for $d=3$, but not for $d=2$, where the disordered phase persists to zero temperature.

For $K=0$ and $J=0$, the model reduces to the Potts and clock models respectively.  For $q=4$ (Figs. 2,3), the expected second-order phase transitions are seen for both Potts and clock models in $d=2$, but in $d=3$ the expected Potts first-order phase transition is narrowly missed in the proximity of a bicritical point.  For $q=5$ (Figs. 2,3), the expected first-order phase transition is not seen for Potts in $d=2$, and the expected Potts first-order phase transition is narrowly missed, in the proximity of a tricritical point, in $d=3$.  Similarly, in $d=2$, the narrow intermediate BKT phase \cite{Polackova} is missed for the clock model.

\section{Conclusion}
The much-used Potts models and clock spin models have been merged into a simple but complex Potts-clock model and solved by renormalization-group theory. The resulting global phase diagram contains a disordered phase and five different ordered phases, namely conventionally ordered ferromagnetic, quadrupolar, and antiferromagnetic phases; algebraically ordered antiferromagnetic and antiquadrupolar phases. These six different phases are separated by first- and second-order phase boundaries, themselves delimited by multicritical points: inverted bicritical, zero-temperature bicritical, tricritical, second-order bifurcation, and zero-temperature highly degenerate multicritical points. A rich sequence of phase diagram topologies thus obtains from a simple model.

\begin{acknowledgments}
Support by the Kadir Has University Doctoral Studies Scholarship Fund and by the Academy of Sciences of Turkey (T\"UBA) are gratefully acknowledged.
\end{acknowledgments}

\end{document}